\documentclass[iop,superscriptaddress]{emulateapj}
\usepackage{amsmath,epsfig,url,natbib}
\usepackage{graphicx,epstopdf,float,color,array}
\usepackage{rotating,multirow}

\shorttitle{WFIRST Redshift Requirements}
\shortauthors{Hemmati et al.}

\submitted{}
\begin{document}
\title{Photometric redshift calibration requirements for WFIRST Weak Lensing Cosmology: Predictions from CANDELS}

\author{Shoubaneh Hemmati\altaffilmark{1}, Peter Capak\altaffilmark{1}, Daniel Masters\altaffilmark{1,2}, Iary Davidzon\altaffilmark{1}, Olivier Dor\`{e}\altaffilmark{2,3}, Jeffrey Kruk\altaffilmark{4}, Bahram Mobasher\altaffilmark{5}, Jason Rhodes\altaffilmark{2,3}, Daniel Scolnic\altaffilmark{6}, Daniel Stern\altaffilmark{2}}
\email{shemmati@caltech.edu}
\altaffiltext{1}{IPAC, California Institute of Technology, 1200 East California Boulevard, Pasadena CA 91125, USA}
\altaffiltext{2}{Jet Propulsion Laboratory, California Institute of Technology, Pasadena, CA 91109, USA}
\altaffiltext{3}{California Institute of Technology, 1201 East California Boulevard, Pasadena, CA 91125, USA}
\altaffiltext{4}{NASA Goddard Space Flight Center, Greenbelt, MD 20771, USA}
\altaffiltext{5}{University of California, Riverside,900 University Ave., Riverside, CA 92521, USA}
\altaffiltext{6}{Kavli Institute for Cosmological Physics, The University of Chicago, Chicago, IL 60637, USA}
\journalinfo{Accepted to the Astrophysical Journal}

\begin{abstract}

In order for Wide-Field Infrared Survey Telescope (\emph{WFIRST}) and other Stage IV dark energy experiments (e.g., Large Synoptic Survey Telescope; LSST, and \emph{Euclid}) to infer cosmological parameters not limited by systematic errors, accurate redshift measurements are needed. This accuracy can be met by using spectroscopic subsamples to calibrate the photometric redshifts for the full sample. In this work we find the minimal number of spectra required for the \emph{WFIRST} weak lensing redshift calibration by employing the Self Organizing Map (SOM) spectroscopic sampling technique. We use galaxies from the Cosmic Assembly Near-infrared Deep Extragalactic Legacy Survey (CANDELS) to build the LSST+\emph{WFIRST} lensing analog sample of $\rm \sim 36$k objects and train the LSST+\emph{WFIRST} SOM. We find that 26\% of the \emph{WFIRST} lensing sample consists of sources fainter than the \emph{Euclid} depth in the optical, 91\% of which live in color cells already occupied by brighter galaxies. We demonstrate the similarity between faint and bright galaxies as well as the feasibility of redshift measurements at different brightness levels. Our results suggest that the spectroscopic sample acquired for calibration to the \emph{Euclid} depth is sufficient for calibrating the majority of the \emph{WFIRST} color-space. For the spectroscopic sample to fully represent the synthetic color-space of \emph{WFIRST}, we recommend obtaining additional spectroscopy of $\sim 0.2-1.2$k new sources in cells occupied by mostly faint galaxies. We argue that either the small area of the CANDELS fields and the small overall sample size or the large photometric errors might be the reason for no/less bright galaxies mapped to these cells. Acquiring the spectra of these sources will confirm the above findings and will enable the comprehensive calibration of the \emph{WFIRST} color-redshift relation.
\end{abstract}


\section{Introduction}

Revealing the nature of the dark energy driving cosmic acceleration and testing general relativity on cosmological scales are essential pieces to complete our understanding of modern cosmology and physics. To achieve these goals, the next generation large cosmology surveys will make precision measurements of the expansion history of the Universe as well as the growth rate of large scale structure using various techniques \citep{Spergel2015}. 

Samples of supernova type Ia (SNIa) constrain cosmological parameters (e.g., the expansion rate of the Universe) by providing measurements of cosmological distances as a function of redshift (e.g., \citealt{Riess1998}, \citealt{Perlmutter1999}). Complementary distance scale measurements can be obtained from baryon acoustic oscillation (BAO) imprints in the power spectrum of cosmic microwave background (CMB) and in the large scale structure (LSS) of galaxies at lower redshifts (e.g., \citealt{Zhan2006}, \citealt{Benitez2009}, \citealt{Aubourg2015}). Weak gravitational lensing of distant galaxies by the gravitational field of matter inhomogeneities in the large scale structure, or cosmic shear, provides another powerful tool for constraining the power spectrum of dark and luminous matter in the Universe (e.g., \citealt{Blandford1991}, \citealt{Blandford1992}). Weak lensing cosmology requires both redshift estimates and shape measurements of statistical samples of galaxies (e.g., \citealt{Hu1999}).

Upcoming stage IV dark energy experiments aimed for 2020s (see \citealt{Albrecht2006}) will improve current measures of distance and cosmic expansion history (with uncertainties $\sim 1-3\%$) as well as matter clustering (with uncertainties $\sim 5-10\%$) to $0.1-0.5\%$ precision, while also extending them to previously unexplored redshift regimes. Careful calibration is required such that the cosmological inferences will not be limited by systematic errors \citep{Spergel2015}. 

Accurate redshifts are needed for all three techniques mentioned above (SNIa, LSS BAO, and weak lensing tomography). While SNIa and BAO studies usually employ a spectroscopic sample, obtaining spectroscopic redshifts for hundreds of millions to billions of faint galaxies needed for weak lensing analysis is not practical. Therefore, highly accurate photometric redshifts, trained and validated using a training sample of spectroscopic data, are required.

Several recent studies (e.g., \citealt{Cunha2012}, \citealt{Newman2015}, \citealt{M15}) have investigated the best spectroscopic sampling strategy in order to train higher-quality, lower-scatter photo-$z$ with less systematic errors for different cosmological surveys. \citet{Carrasco2013} showed that random selection of galaxies to create a spectroscopic training sample is not optimal. Recent work has suggested spatial cross-correlation-based techniques relating the photometric redshifts with a reference spectroscopic sample as a solution (e.g., \citealt{Newman2008}, \citealt{Newman2015}). These techniques also require a large spectroscopic sample. However, their main advantage is that the spectroscopic sample does not need to be representative of different galaxy types (i.e. bright emission line galaxies which are easy spectroscopic targets can be used for calibration). 

\begin{figure} 
\centering
  \includegraphics[trim=0cm 0cm 0cm 0cm, clip,width=0.5\textwidth] {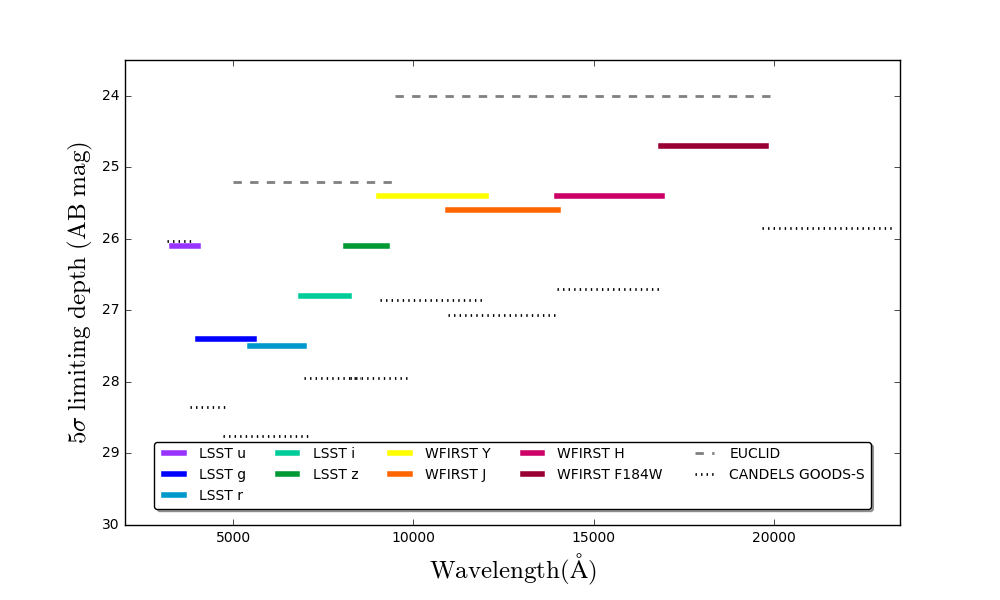}
\caption{Expected 5$\sigma$ limiting magnitudes of LSST and the \emph{WFIRST} high latitude survey (HLS) filters for galaxies ($\rm r_{1/2}=0.3"$) plotted as solid colored lines. We use deeper or similar depth photometry in the same wavelength range as probed with LSST+\emph{WFIRST} from CANDELS catalogs in five fields to estimate LSST+\emph{WFIRST} photometry. 5$\sigma$ limiting AB magnitude of CANDELS GOODS-S filters for galaxies ($\rm r_{1/2}=0.3"$) are plotted with dotted black lines (CANDELS filters used are listed in Table \ref{tbl:filters}). Euclid's expected riz as well as NIR 5$\sigma$ depths are also over plotted as dashed gray lines for comparison.}
\label{fig:filters}
\end{figure}

A completely data-driven technique of selecting optimal spectroscopic samples to meet the cosmological requirements was introduced by Masters et al. (2015; hereafter M15). This technique uses a machine learning algorithm called the Self Organizing Map (SOM) (Kohonen 1982, 1990) to reduce the multi-dimensional color-space of galaxies defined by a photometric survey to two dimensions (hence maps). This empirical color mapping method allows us to focus our spectroscopic efforts on under-sampled regions of galaxy parameter space. M15 explored different SOM-based targeting strategies and estimated the required spectroscopy for the \emph{Euclid} mission (\citealt{Laureijs2011}). This approach is the basis of a large, ongoing spectroscopic program, the Complete Calibration of the Color-Redshift Relation (C3R2) survey, designed to calibrate the color-redshift relation to the \emph{Euclid} depth (\citealp{Masters2017}, Masters et al. in prep). Recently, \cite{Sanchez2019} presented a framework based on hierarchical Bayesian model to infer redshift distributions from the combination of galaxy colors and clustering information. SOMs can also be used to define galaxy phenotypes based on these Bayesian schemes, specifically where more limited and noisy colors exist for galaxies.

In this paper, as part of the High Latitude Survey (HLS- \citealt{Dore2018}) science investigation team, we extend the previous analysis of M15 to estimate the additional spectroscopic sample required to meet the Wide-Field Infrared Survey Telescope (WFIRST) cosmological requirements. \emph{WFIRST} is a NASA flagship mission using a 2.4-meter telescope to provide measurements of the expansion history  of the universe and growth of structure to better than 1$\%$ (\citealt{Spergel2015}). For weak lensing analysis, the \emph{WFIRST} HLS is currently planing to image $\rm 2227 \ deg^{2}$ in four near infrared (near-IR) bands ($Y$, $J$, $H$, and $F184$) spanning the range from $0.92\!-\!2.00\,\mu\mathrm{m}$ to magnitudes $\rm 25.8\!-\!26.7$ (depending on band), significantly fainter compared to the near-IR depths in the \emph{Euclid} survey ($\sim 24.5$). The near IR filters alone are not sufficient for precise photo-$z$ estimation. Multi-band optical observations need to be combined with \emph{WFIRST} to fulfill the redshift requirements. Such observations will be available through  the Hyper Suprime-Cam (HSC) on Subaru or by the Large Synoptic Survey Telescope (LSST).

In \S 2 we simulate a data-driven photometric catalog using deep observations from the Cosmic Assembly Near-IR Deep Extragalactic Legacy Survey (CANDELS; \citealt{Grogin2011}, \citealt{Koekemoer2011}) to replicate the LSST+\emph{WFIRST} lensing sample. In \S 3 we briefly review the SOM technique, train the SOM with the LSST+\emph{WFIRST} analog catalog and test for its accuracy in representing the data. We check for the effects of cosmic variance in \S 4 and address the additional spectroscopy needed to meet \emph{WFIRST} cosmology requirement in \S 5. \S 6 summarizes the results of this work and discusses sources of uncertainty. Throughout this paper all magnitudes are expressed in AB system (\citealt{Oke1983}) and we use the concordance $\Lambda$CDM cosmology with $H_{0}=70 \ \rm km \ s^{-1}\  Mpc^{-1}$, $\Omega _{M}= 0.3$, and $\Omega _{\Lambda} = 0.7$.

\begin{figure} 
\centering
  \includegraphics[trim=0cm 0cm 0cm 0cm, clip,width=0.50\textwidth] {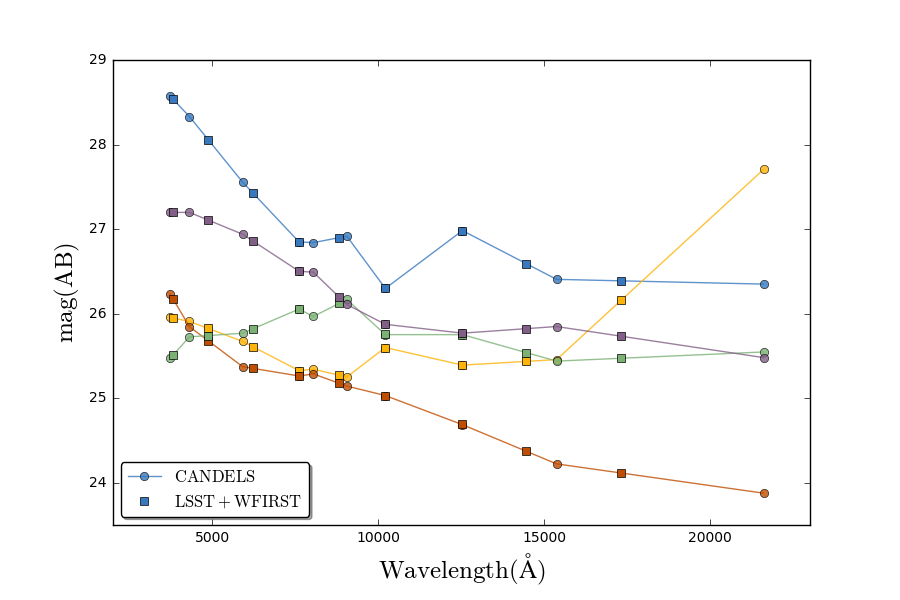}
\caption{LSST and \emph{WFIRST} measured photometry for five sample galaxies from CANDELS GOODS-S (squares). Solid lines and circles show the CANDELS broadband filters used for the linear interpolation. Each color represents a sample galaxy.}
\label{fig:interpol}
\end{figure}

\section{\emph{WFIRST} $+$ LSST Weak Lensing Sample \label{sec:data}}

\subsection{CANDELS Galaxy Sample}

\begin{table*}
\footnotesize
\centering
\caption{CANDELS filters used in each field to create the LSST+\emph{WFIRST} catalog}
\begin{tabular}{*{15}{c}}
\hline
\hline
\\
Field&&&&&&& Filters\footnotemark&&\\
\\
\hline
\\
GOODS-S& $U_{VIMOS}$& $\rm F435W$& $\rm F606W$&$\rm F775W$ & $\rm F814W$ & $\rm F850lp$ & $\rm F098W$& $\rm F105W$ & $\rm F125W$ &$\rm F160W$&$Ks_{HAWK-I}$\\
\\
GOODS-N&$U_{KPNO}$& $\rm F435W$& $\rm F606W$&$\rm F775W$ & $\rm F814W$ & $\rm F850lp$ &$\rm F105W$ & $\rm F125W$ &$\rm F160W$& $Ks_{CFHT}$\\
\\
EGS& $ U_{CFHT}$& $g_{CFHT}$& $\rm F606W$ &$ r_{CFHT}$& $i_{CFHT}$& $\rm F814W$ &$ z_{CFHT}$ & $\rm F125W$&$\rm F160W$&$Ks_{CFHT}$\\
\\
UDS& $ U_{CFHT}$& $ B_{subaru}$&$\rm F606W$ &$ Rc_{subaru}$& $i_{subaru}$&$ F814W$ &$ z_{subaru}$ & $Y_{HAWK-I}$&$\rm F125W$&$\rm F160W$&$ Ks_{HAWK-I}$\\
\\
COSMOS&$ U_{CFHT}$& $ B_{subaru}$& $\rm F606W$&$ r_{subaru}$&$ i_{CFHT}$&$\rm F814W$ &$ z_{CFHT}$ & $ Y_{UVISTA}$&$\rm F125W$&$\rm F160W$& $ Ks_{UVISTA}$\\
\\
\hline
\end{tabular}
\footnotetext{\footnotesize Refer to CANDELS catalog papers for detailed description of observations in each filter}
\label{tbl:filters}
\end{table*}

CANDELS photometric catalogs are the optimum choice for testing \emph{WFIRST} HLS redshift requirements as they are H-band selected and provide deeper multi-band observations compared to WFIRST. CANDELS obtained very deep near-IR observations of five different fields using the {\it Hubble Space Telescope} (\emph{HST}). These observations are homogeneously combined with the wealth of ancillary space- and ground-based data available from UV to X-ray.

\begin{figure} 
\centering
  \includegraphics[trim=0cm 0cm 0cm 0cm, clip,width=0.50\textwidth] {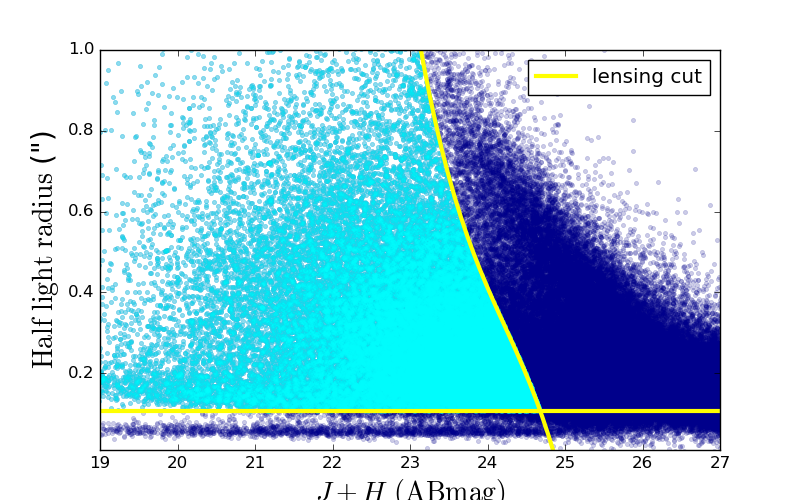}
\caption{Subsample required for lensing analysis (aqua dots) selected from the LSST+\emph{WFIRST} simulated photometric catalog (dark blue dots) by a cut based on size and \emph{J+H} AB magnitude (yellow solid lines) which requires the galaxy to be resolved, have a high S/N ($\rm >18$) and small ellipticity uncertainty ($\rm <0.2$).}
\label{fig:lenscut}
\end{figure}

Source detection in all CANDELS catalogs was conducted in the Wide Field Camera 3 (WFC3) F160W band (at 1.6 $\mu$m) and photometry is generated using the Template FITting algorithm (TFIT; \citealt{Laidler2007}). Details of CANDELS observations and photometric catalogs for each of the five fields used in this study can be found in the published catalog paper for each field; the GOODS South field \citep{Guo2013}, the UDS field \citep{Galametz2013}, the COSMOS field \citep{Nayyeri2017}, the Extended Groth Strip field (EGS; \citealt{Stefanon2017}), and the GOODS North field (Barro et al. in prep). The published photometric redshifts by the CANDELS team, which we use in this work are based on combining results from multiple teams each using a different combination of photometric redshift code, library of template spectral energy distributions (SEDs), and priors (see e.g., \citealt{Dahlen2013}). These measurements have rms of $\sim 0.03$ with an outlier fraction of at most $\sim 3 \%$, measured from a spectroscopic sample. \cite{Dahlen2013} found a strong magnitude dependence in the accuracy of photometric redshifts and hence the reported uncertainties might be slightly underestimated for spectroscopic samples biased towards brighter objects. In this work, where we use a subsample of CANDELS galaxies suitable for lensing analysis (discussed in section 2.3) the depth of the spectroscopic sample is comparable to our targets and hence the level of uncertainty is still below what is expected for individual photoz precision for weak lensing analysis in stage Iv cosmology experiments (e.g., \citealt{Bordoloi2010}).

\begin{figure} 
\centering
  \includegraphics[trim=0cm 0cm 0cm 0cm, clip,width=0.50\textwidth] {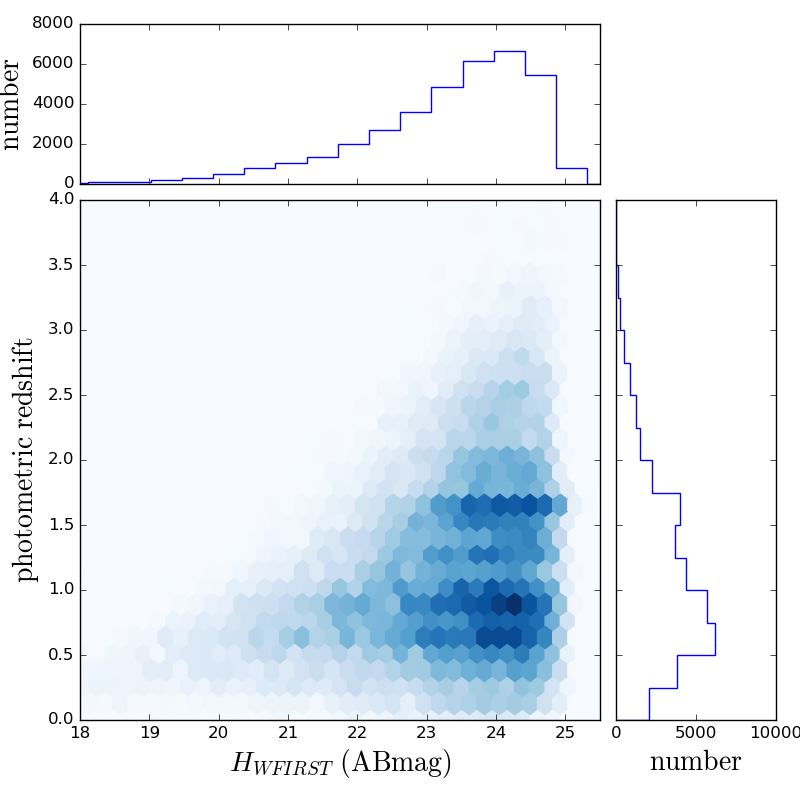}
\caption{Density of the simulated lensing sample (a total of 36,612 galaxies selected from all five CANDELS fields which meet the lensing cut criteria) in the photometric redshift vs. $H_{WFIRST}$ magnitude plane plotted as blue hexagons with darker colors representing higher densities of objects.}
\label{fig:sample}
\end{figure}

The total area covered by the CANDELS (wide) observations is $\rm \sim 0.2\ deg^{2}$ reaching $\rm 5 \sigma$ limiting magnitude depth of $\rm \sim 26.5$ in the WFC3 F160W observations, more than a magnitude deeper than the planned depth for \emph{WFIRST} (see Figure \ref{fig:filters}). The small area covered by the CANDELS fields compared to the \emph{WFIRST} HLS can raise concerns about the effects of cosmic variance in the analysis. We address this issue briefly in \S 4.  

\begin{figure*} 
\centering
  \includegraphics[trim=0cm 0cm 0cm 0cm, clip,width=0.98\textwidth] {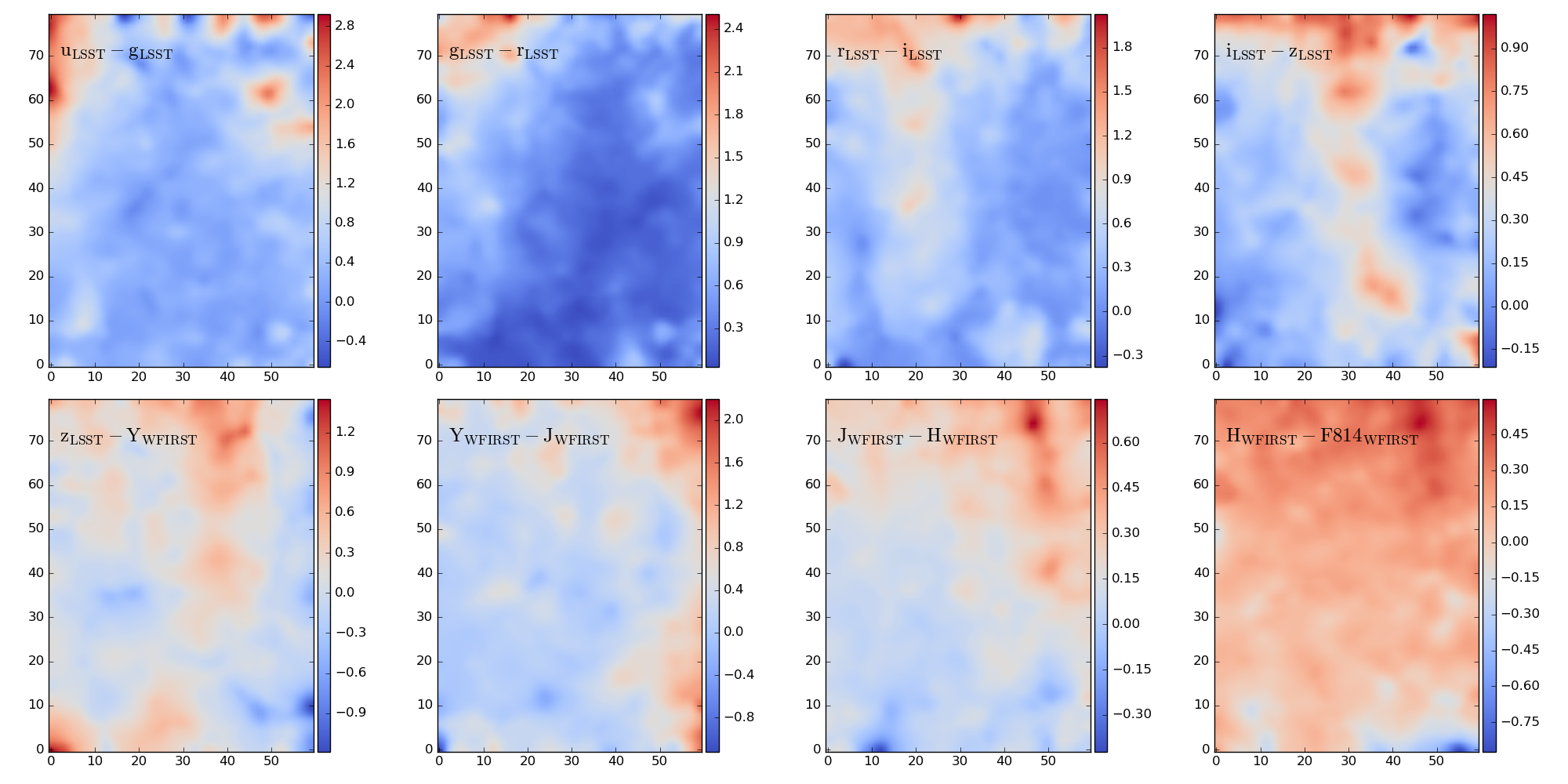}
\caption{LSST and \emph{WFIRST} colors of the trained SOM at each cell from top-left to bottom-right color-coded by: $\rm u_{LSST}-g_{LSST},\ g_{LSST}-r_{LSST},\ r_{LSST}-i_{LSST},\ i_{LSST}-z_{LSST},\ z_{LSST}-y_{WFIRST},\ y_{WFIRST}-j_{WFIRST},\ j_{WFIRST}-H_{WFIRST},\ and\ H_{WFIRST}-F184W_{WFIRST}$. SOM is selected to be a mesh of $\rm 80\times 60$ cells. The axes are arbitrary and each position on the two dimensional map points to a position in the 8 dimensional color space.}
\label{fig:input_som}
\end{figure*}

\subsection{From CANDELS to LSST+\emph{WFIRST} Photometry}
To estimate the photometry of CANDELS galaxies in each of the nine LSST and \emph{WFIRST} filters, we use linear interpolation of the photometry in the deepest CANDELS filters that straddle the LSST and \emph{WFIRST} filters in each field. Figure \ref{fig:filters} shows the LSST and \emph{WFIRST} filters and their expected $\rm 5 \sigma$ limiting depths (HLS expected depths are shown for \emph{WFIRST}) in comparison to CANDELS GOODS-S filters used to estimate the photometry in the new filters. Table \ref{tbl:filters} lists the CANDELS filters used in each field for these measurements. Figure \ref{fig:interpol} shows five sample SEDs and the estimated LSST+\emph{WFIRST} photometry from linear interpolation. We extensively tested this method of linear interpolation for estimating broad-band photometry. The interpolation technique reproduces more realistic color distributions compared to fitting galaxies with theoretical (or empirical) model SEDs and convolving best fitted model SEDs with the broad-band filters. This technique is model independent and data-driven therefore lacks SED modeling uncertainties. Furthermore, the CANDELS bands are well matched to the WFIRST/LSST bands and deeper, so the linear interpolation introduces minimal errors with respect to the true photometry. 

\subsection{\emph{WFIRST} Lensing Analog Sample}
As the main purpose of this study is to find the spectroscopic sample requirement for weak-lensing, we apply a cut to the photometric catalog to exclude galaxies which will not benefit the lensing analysis (based on private communication with C. Hirata). This cut removes galaxies whose shape distortion measurements will not be accurate enough for weak lensing analysis. Figure \ref{fig:lenscut} shows this lensing criteria applied to the sample based on the galaxy FWHM in the F160W filter and the average \emph{J+H} magnitude. This criteria is driven from requiring the galaxy to be resolved, the average \emph{J+H} flux have $\rm S/N>18$ and the ellipticity uncertainty $\rm <0.2$. The limiting flux depends on the galaxy size. The cut applied is deliberately inclusive, and if specific regions of color-magnitude-size space are found to be difficult for lensing analysis and shape measurement, those could be excluded from lensing analysis later. The lensing cut applied here, also excludes the very faint galaxies that exist in deep CANDELS data but will not be in the \emph{WFIRST} due to magnitude limits.

The final photometric catalog of the lensing sample consists of 36,612 galaxies with estimates of their photometry in LSST optical and \emph{WFIRST} near IR filters, as well as their photometric and spectroscopic (when available) redshifts, and physical sizes. Figure \ref{fig:sample} shows the distribution of this sample in \emph{H} band and redshift, with the majority of sources at $H_{WFIRST}\sim 24$ mag and $z\sim 1$.
%%%%%%%%%%%%%%%%%%%%%%%%%%%%%%%%%%%%%%%%%%%%%%%%%%%%%%%%%%%%%%%%%%%%%%%%%%%%%%%%%%%%%%%%%%%%%%%%%%%%%%%
\section{2D Map of the \emph{WFIRST} color space \label{sec:som}}

SOMs offer an optimized and efficient way to target sources for spectroscopy from the less explored regions of galaxy color-space and provide the means to calibrate the photometric redshifts using the acquired spectroscopic sample. In short, SOMs introduced in the 1980s by Kohonen, are a class of unsupervised artificial neural networks which reduce dimensionality of a multi-dimensional parameter space (color-space of galaxies in this case), while preserving the topology in the parameter space. In other words similar objects in multi-dimensional parameter space remain neighboring on two dimensional grids (maps). Therefore SOMs are also an optimized way of visualizing multi-dimensional parameter space.

\begin{figure*}
\centering
  \includegraphics[trim=0cm 0cm 0cm 0cm, clip,width=0.98\textwidth] {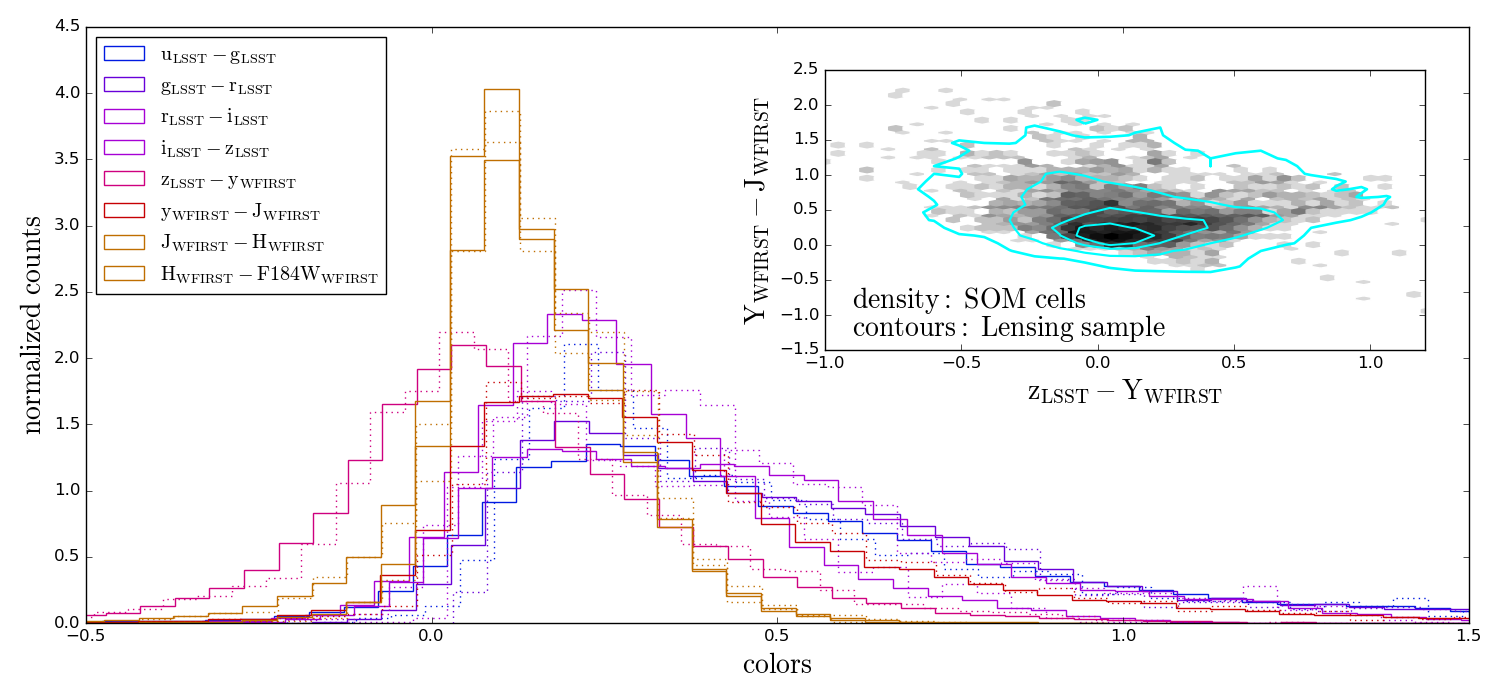}
\caption{Comparison of normalized distribution of colors of galaxies in the input lensing catalog (solid) and colors of SOM cells (dashed) in the eight colors used for training the SOM. Inset shows an example color-color distribution ($Y_{WFIRST}-J_{WFIRST}$ vs. $\rm z_{LSST}-Y_{WFIRST}$) of the SOM cells with gray density plot and the aqua contours represent the lensing sample distribution with 10, 100, 500, 1000 levels. The similarity between the distributions proves that the SOM is trained well and that the SOM cells represent the training data accurately. }
\label{fig:test_som1}
\end{figure*}

\begin{figure*} 
\centering
  \includegraphics[trim=0cm 0cm 0cm 0cm, clip,width=0.98\textwidth] {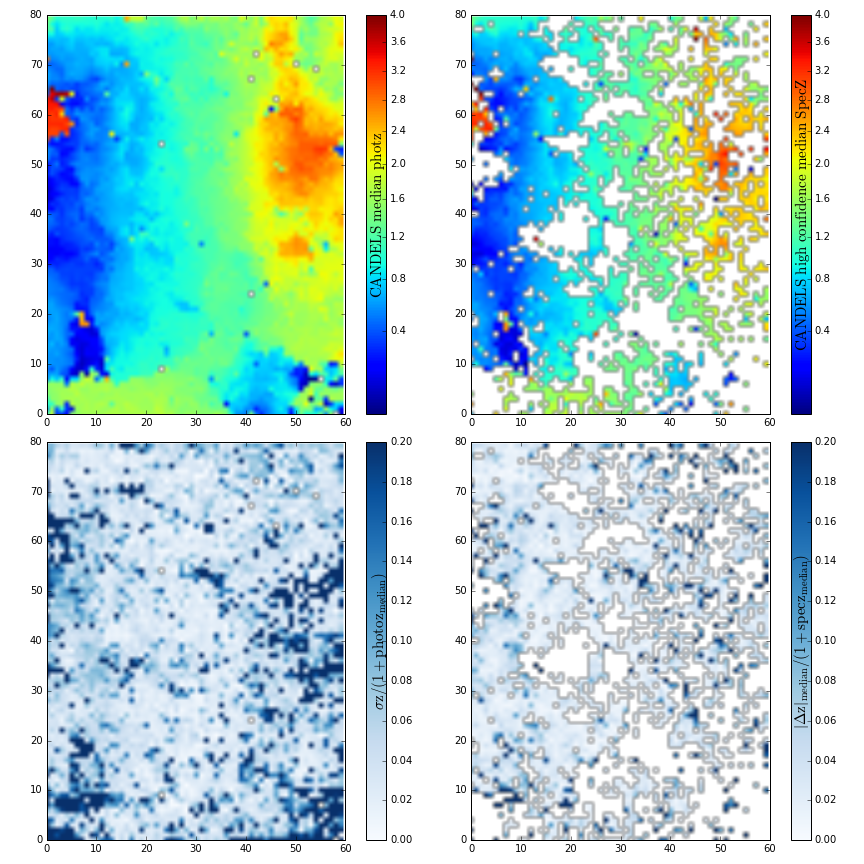}
\caption{Distribution of the median photometric redshifts (top-left), median of high confidence spectroscopic redshifts (top-right), the full scatter $\sigma_{F}$ of photometric redshifts defined as $\sigma photoz/(1+z_{median})$ (bottom-left), and  the redshift bias, defined as $\rm median |specz - photoz| /(1+specz)$ (bottom-right) on the SOM.}
\label{fig:phot-z-spec-z}
\end{figure*}

\begin{figure*} 
\centering
  \includegraphics[trim=0cm 0cm 0cm 0cm, clip,width=0.98 \textwidth] {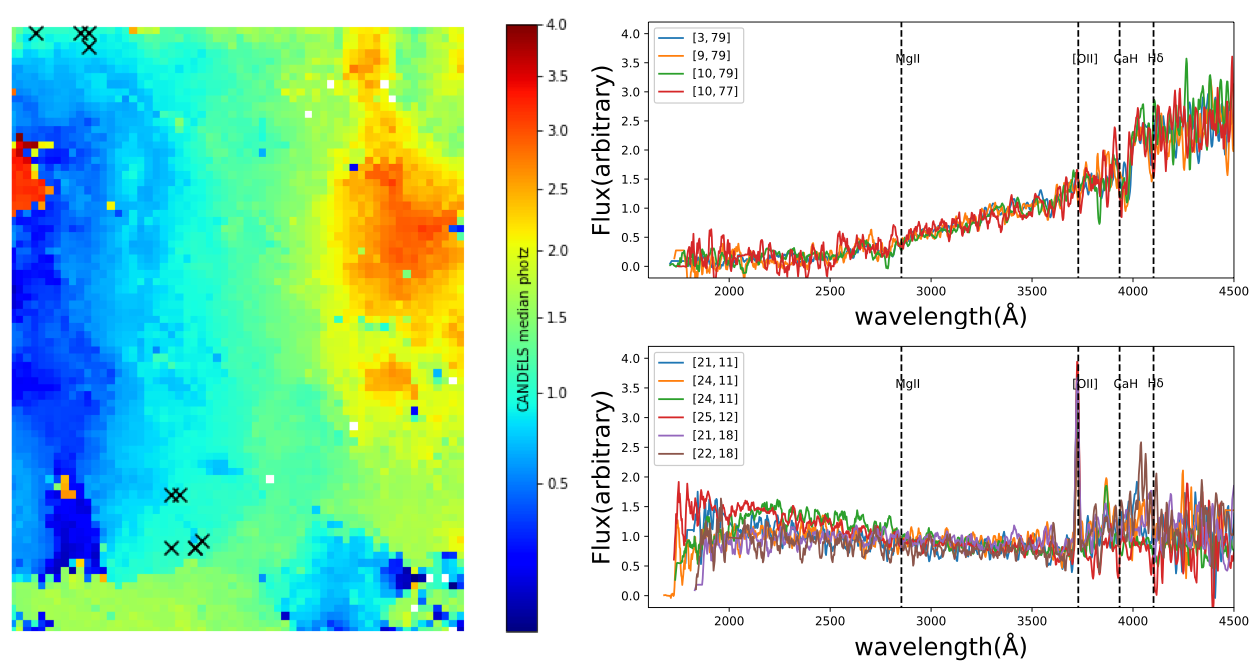}
\caption{Ten $z \sim 1$ galaxies from the VVDS which have high quality 1D spectra are mapped to the LSST+\emph{WFIRST} SOM. On left panel we show where these ten galaxies sit on the LSST+\emph{WFIRST} SOM (color-coded by redshift) with black crosses. Right panels top and bottom respectively, show the normalized 1D spectra of the sources mapped to the top and bottom regions of the SOM. From this it can be clearly seen how SOM is effectively grouping similar galaxies closer together.}
\label{fig:vvds_spec}
\end{figure*}

\subsection{Training the SOM with WFIRST-Analog Data}
In this work, we use the {\sc sompy} package, a \texttt{python} library for self organizing maps. We assume a rectangular topology and $80\times 60$ cells. This grid size is optimized to represent the color-space of the training sample (in section 5 we discuss the effects of choosing different grid sizes). Similar to most other artificial neural networks there is a training phase and a mapping phase. We use the LSST+\emph{WFIRST} colors of galaxies in the lensing sample ($ u_{LSST}-g_{LSST},\ g_{LSST}-r_{LSST},\ r_{LSST}-i_{LSST},\ i_{LSST}-z_{LSST},\ z_{LSST}-Y_{\emph{WFIRST}},\ Y_{\emph{WFIRST}}-J_{\emph{WFIRST}},\ J_{\emph{WFIRST}}-H_{\emph{WFIRST}},\ and\ H_{\emph{WFIRST}} F184_{\emph{WFIRST}}$) to train the SOM. Each cell in the SOM will have a weight vector with the dimensions of the input training data (eight dimensions here). We assign the initial weights by Principle Component Analysis (PCA). During the competitive training phase, weight vectors adapt themselves to represent the distribution and topology of the input data. This is done by finding the best matching unit (BMU) of the SOM for each data point in the training set and bringing the weight vector of the BMU, and its neighbors, closer to the training data point. The magnitude and radius of the change decrease over time until the SOM represents the data well. {\sc SOMPY} does the training in batch mode, where in each epoch the SOM is exposed to the entire training sample. Batch-mode training is generally expected to be quicker and result in a more stable network. Figure \ref{fig:input_som} represents our trained SOM colored by each component (color) of the final weight vector at each cell. Each position on the two dimensional map points to a position in the eight dimensional color space.

The final quantization error of the SOM, which is defined as the mean of the Euclidean distances of all training data to their BMUs, is 0.81. To further verify that the SOM is trained properly and that the weight vectors of the neurons (cells) adequately represent the training data, we compare the distribution of SOM cell colors to lensing sample colors in Figure \ref{fig:test_som1}. The identical range and shapes of the distributions confirm qualitatively that the SOM cells represent the training data.

\subsection{Mapping \emph{WFIRST} Colors to Redshift}

Figure \ref{fig:phot-z-spec-z} presents redshift information of the lensing galaxy sample on the SOM. We map galaxies in the LSST+\emph{WFIRST} analog catalog back to the trained SOM and color code the SOM by the median photometric redshifts of galaxies in each cell (top-left panel), where the photometric redshifts are measured by the CANDELS team (\citealt{Dahlen2013}, \citealt{Nayyeri2017}, \citealt{Stefanon2017}, Barro et al. in prep). The smoothness of the photometric redshift map indicates that the combined colors in LSST and \emph{WFIRST} filter sets are adequate for photometric redshift estimation. A redshift uncertainty map defined as $\frac{\sigma_{z}}{1+photoz_{median}}$ where $\sigma_{z}$ is the standard deviation of redshifts of galaxies in each cell is shown in the bottom-left panel of Figure \ref{fig:phot-z-spec-z}. The average uncertainty per cell is small, of the order of 0.04, while larger uncertainties ($\sim 0.2$) are found on the boundaries between high and low redshift regions of the SOM. High confidence spectroscopic redshifts in CANDELS, where available, are also shown on the SOM (top-right panel) and cover $\sim 57 \% $ of the color-space. These are from public spectroscopic redshifts available in the CANDELS fields, i.e. the CANDELS public compilation of spectroscopic redshifts (by N. Hathi), the MOSDEF public spectroscopic redshift catalog (\citealt{Kriek2015}), and the 3D-HST (\citealt{Brammer2012}) catalog of grism spectroscopy. While more than half of the SOM cells are occupied by at least one high quality spectroscopic redshift, this corresponds to only $\sim 20\%$ of the sample (i.e. 7453 spectra). Visual comparison with the photometric redshift map shows good agreement between the photometric and spectroscopic redshifts. As discussed thoroughly in M15, using the SOM technique in calibrating the redshifts has the advantage of showing whether the training sample is representative of the color distribution of galaxies in a survey. The high confidence spectroscopic redshifts in each cell are necessary to calibrate the redshift map and hence the color-redshift relation to the accuracy needed by cosmology. The redshift bias parameter is also estimated (bottom-right panel) and is defined as ($\frac{ |\Delta z|_{median}}{1+specz_{median}}$), where $\rm |\Delta z|_{median} \equiv median (photoz - specz_{median})$ and has lower values compared to the photometric uncertainty map with median of $\sim 0.03$. Most of the higher uncertainty cells seen in the photometric redshift uncertainty map (e.g., lower right and left corners of Figure \ref{fig:phot-z-spec-z}) coincide with those where high confidence spectroscopic data are missing. This is due to biases in spectroscopic samples, not covering all the color-space, and larger photometric uncertainties leading to less certain photometric redshifts in these regions. 

\subsection{Beyond Redshifts and Broad-band Photometry}

Thus far we have trained and tested the LSST+\emph{WFIRST} SOM using the eight optical and near-IR colors of the lensing analog sample. In this section, we go one step further by analyzing the high-resolution spectra of a sample of galaxies across the SOM. We draw galaxies from the VIMOS VLT Deep Survey (VVDS) which is a comprehensive deep galaxy spectroscopic redshift survey conducted by the VIMOS collaboration with the VIMOS multi-slit spectrograph at the ESO-VLT (\citealt{LeFevre2013}). The VVDS spectra also span a broad range of magnitudes with no cut on object properties other than their $I-$band magnitude. As a result we can test how rapidly spectral properties change with position on the SOM.

Figure \ref{fig:vvds_spec} shows some of the VVDS $z\sim 1$ galaxies with high confidence spectroscopic redshifts that grouped on two distant regions of the SOM. These galaxies lie in the $z \sim 1$ region as shown from the underlying color of the SOM (left panel). We smooth and normalize the 1D spectra of these sources and shift them to $z=0$ to compare. Spectra of galaxies mapped to upper and lower regions of the SOM are shown in the right panel of Figure \ref{fig:vvds_spec}. This figure shows that the SOM, trained by broadband colors, can also statistically group galaxies with similar spectral features closer together. This is akin to previous works using broadband photometry to estimate emission line ratios in galaxies (e.g., \citealt{Faisst2016}). Here, for instance, the galaxies in the lower parts of the SOM show strong nebular emission lines (e.g., [OII]) while the upper ones do not. This test illustrates the SOM classifies galaxies by their intrinsic spectral shape, and that high-resolution spectral features are captured by the SOM to some extent.  The amount of high-resolution information is clearly limited by the photometric precision and spectral resolution of the filter set, but is sufficient to capture some key features. 

\begin{figure} 
\centering
  \includegraphics[trim=0cm 0cm 0cm 0cm, clip,width=0.50\textwidth] {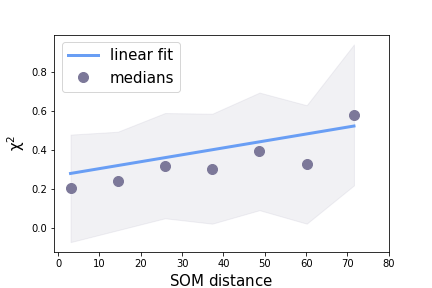}
\caption{$\chi ^{2}$ difference between VVDS spectra vs. their distance on the SOM. We mapped multi-band photometric SEDs of all sources with high quality spectra from the VVDS, to our trained SOM and measured distances of 2000 pairs of randomly selected galaxies with $dz=1$. Blue solid line is the linear fit to the data. Gray circles and shaded region represent the running median and one sigma dispersion, respectively. }
\label{fig:x2dis}
\end{figure}
Detailed quantification of spectral similarity is not straight forward and is beyond the scope of the current paper. Here we present a simple $\chi^{2}$ analysis and leave a more detailed exploration of systematics and recovery of spectral features with simulated spectra to a future paper (Hemmati et al. in preparation). The $\chi^{2}$ difference between the spectra shown in the right panel of Figure \ref{fig:vvds_spec}, is of the order $\sim 0.1$ and $\sim 0.15$ among spectra in top and bottom panels respectively. The $\chi ^{2}$ increases to more than 1 when spectra from top and bottom panels are compared. In Figure\ref{fig:x2dis} we map all 460 galaxies with high quality spectroscopic redshifts from VVDS to the SOM (redshift range  $0<z<3.5$). Then we randomly pick two thousand pairs of galaxies, where the second galaxy had to be within a $dz=0.1$ from the first random selection. We measured their distance on the SOM and the $\chi ^{2}$ difference between their smoothed spectra. The overall trend of decreasing similarity between spectra with increasing distance on the SOM can be clearly seen from this exercise. However, we note that while $\chi ^{2}$ does present a measure of similarity it is not the best way to distinguish similar features such as emission lines between noisy spectra. Also, while distance on the SOM is an overall measure of how different shapes of galaxy SEDs are, a more careful analysis would be to use derivative based clustering on the SOM, as distances on SOMs are not preserved after projection from multi-dimension.

\begin{figure*} 
\centering
  \includegraphics[trim=0cm 0cm 0cm 0cm, clip,width=0.98\textwidth] {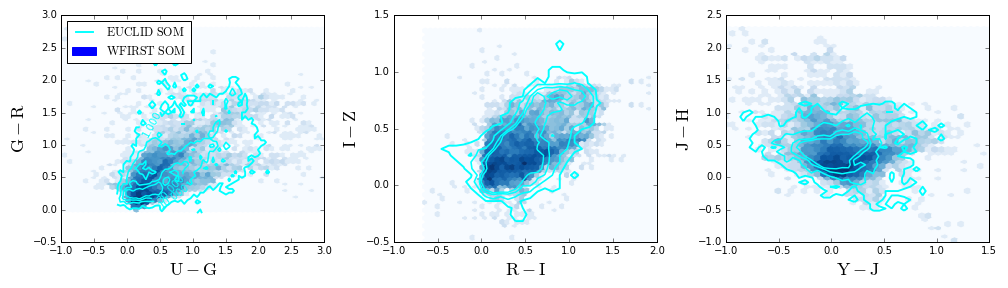}
\caption{Comparison of color distribution of SOM cells trained with \emph{WFIRST} lensing sample (blue density) and \emph{Euclid} sample (aqua contours, with [1,10,30,100] contour levels). The \emph{Euclid} sample is taken from the combination of COSMOS, SXDS, and VVDS surveys covering a total of 3.8 $\rm deg^{2}$, 19 times larger than the area covered by the five CANDELS fields. }
\label{fig:euclidsom}
\end{figure*}

\section{Cosmic Variance}

The CANDELS survey area is much smaller than what would be covered by the \emph{WFIRST} HLS, so it is important to determine the effects of cosmic variance on redshift calibration. Despite this concern, the depth of multi-band observations by CANDELS makes it the most favorable option to simulate the photometry of \emph{WFIRST} galaxies. The effect of cosmic variance in measurement error has been claimed to be an order of magnitude larger than Poisson errors for determining redshift distributions (e.g., \citealt{Newman2002}, \citealt{Cunha2012}). \cite{Newman2015} suggested that a minimum of 10-20 fields of $\sim 20$ arcmin diameter each (to cover multiple correlation lengths) is needed in order to meet the requirements for photometric redshift calibration. Here, the CANDELS fields cover only 5 widely-separated fields in the sky covering a total of $\sim 0.2\,\mathrm{deg}^2$. 

We only briefly explore the effect of the small area in this work and defer the full quantitative analysis to a future paper. We visually compare the SOM trained by our \emph{WFIRST} lensing sample to the \emph{Euclid} SOM (\citealt{Masters2017}) which is trained using the combination of COSMOS (\citealt{Scoville2007}), SXDS (\citealt{Furusawa2008}), and VVDS (\citealt{LeFevre2005}) surveys covering $3.8\,\mathrm{deg}^2$ and is therefore 19 times larger. Figure \ref{fig:euclidsom} presents this comparison on three color-color plots with the density of colors of the \emph{WFIRST} lensing sample SOM shown with shades of blue and \emph{Euclid} SOM over-plotted with aqua contours. The range and shapes of these distributions are visually identical. This indicates that increasing the area by $\sim 20$ times does not expand the color-space, but rather fills the gaps between the already covered color-space. 

Photometric redshifts measured from multi-band SEDs need spectroscopic observations to calibrate the systematics in redshift measurements due to lack of precision in low resolution SEDs. Selecting spectroscopic samples of galaxies randomly or based on environment for redshift calibration suffers from cosmic variance, as by definition they are do not cover the whole range of possible SED shapes. Selecting the spectroscopic calibration sample systematically from the well-occupied color-space instead should not suffer as much from the loss of different types of galaxies. However, we note that the quantification of cosmic variance in redshift calibration with SOMs for weak lensing is critical and can not be done with these small observed data sets. Capak et al.~(in prep) explore the effect of cosmic variance in more depth, exploiting large cosmological simulations of galaxies, comparable in size to that of the \emph{WFIRST} HLS.

%%%%%%%%%%%%%%%%%%%%%%%%%%%%%%%%%%%%%%%%%%%%%%%%%%%%%%%%%%%%%%%%%%%%%%%%%%%%%%%%%%%%%%%%%%%%%%%%%%%%%%%%%%%%%%
\section{Optimal Sampling Technique to Meet Weak Lensing Redshift Requirement \label{sec:sampling}}

Different spectroscopic sampling strategies from the SOM were explored in M15 to calibrate the $\langle z \rangle$ of the tomographic redshift bins to the required level for weak lensing cosmology ($\Delta \langle z \rangle < 0.002 (1+\langle z \rangle)$; see e.g., \citealt{Bordoloi2010}, \citealt{Kitching2008}). M15 showed that with simplifying assumptions, if $\sigma_{\langle z \rangle}$ in the SOM cells are of the order of $\sim 0.05$, with $\sim 600$ color cells (c) in each tomographic bin and assuming that the mean redshift of each cell is known ($\sim 1$ spectra per cell), the calibration requirement can be met (see M15 for details, in short $\Delta \langle z \rangle \simeq \sigma_{\langle z \rangle} / \sqrt{c}$). In M15, this meant a total of $\sim 10,000-15,000$ spectra, much lower compared to estimates of direct calibration through random sampling. The gain in statistical precision from the SOM method compared to direct sampling is attributed to the systematic way the full color-space is sampled.

\begin{figure*} 
\centering
  \includegraphics[trim=0cm 0cm 0cm 0cm, clip,width=0.98\textwidth] {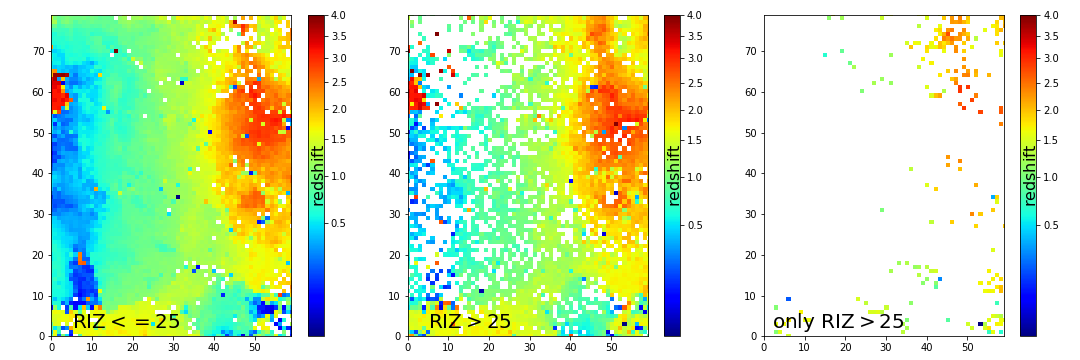}
\caption{Bright ($\rm riz<25$; \emph{Euclid} depth) and faint ($\rm riz>25$) galaxies in \emph{WFIRST} lensing sample are mapped to the SOM color coded by median redshifts (shown on left and middle panels). More than 95\% of the SOM cells contain at least one bright galaxy, $\sim 71 \%$ of the SOM cells contain at least one faint object, and only $\sim 4\%$ of cells contain only faint galaxies (right panel).}
\label{fig:euclid}
\end{figure*}

\begin{figure} 
\centering
  \includegraphics[trim=0cm 0cm 0cm 0cm, clip,width=0.5\textwidth] {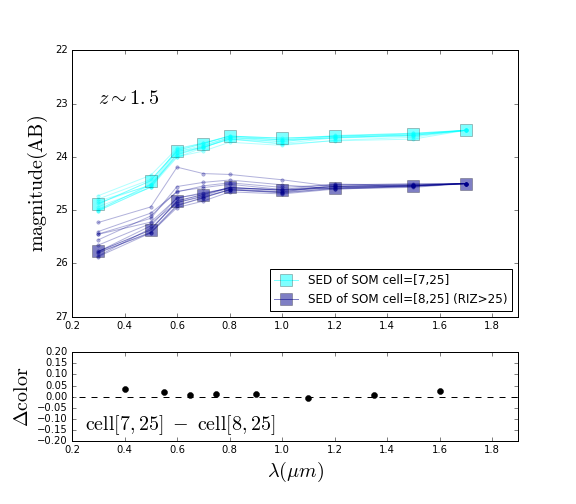}
\caption{SED of a SOM cell $[8,25]$ containing only faint ($riz>25$) objects shown with dark blue squares and the SED of objects mapped to this cell presented as dark blue dots and solid lines is compared to its neighboring cell $[7,25]$ containing both bright and faint objects shown with aqua squares and the SED of objects mapped to this cell presented as aqua dots and solid lines. The F184W magnitude of SOM cell SEDs are assigned manually to demonstrate bright vs. faint, and the rest are from the colors of the cell. The bottom panel shows color differences between the two cells, with the x axis being the average wavelength between the two filters of each color.}
\label{fig:sed_faint}
\end{figure}

Average redshift uncertainties in the LSST+\emph{WFIRST} SOM provided here are smaller compared to the estimates in M15 ($\sim$ 0.03 vs. $\sim$ 0.05) due to higher S/N in \emph{WFIRST} observations. The average number of cells needed in each tomographic bin to meet the calibration requirement (based on simplified assumptions mentioned above) is then reduced to $\sim 200$. Given an $\rm 80\times 60$ grid for the SOM, there can be 24 tomographic bins of this size and to fill the SOM with at least one spectroscopic redshift per cell, $\sim 5 $k spectra would be needed. This does not correspond to $5 $k new spectra, as many CANDELS galaxies have prior spectroscopic observations (see Figure \ref{fig:phot-z-spec-z}). Also there are galaxies with almost identical SED shapes in other fields of the sky which we can use to calibrate this SOM (discussed in following subsections).
 
 It is important to note that, the number of spectra needed for calibration is independent of the initial size of the rectangular grid chosen for the SOM. Having a smaller rectangular grid, will increase the average redshift scatter in SOM cells. This is because more distant SEDs are grouped together in a cell as compared to having a larger grid, which leads to having a higher redshift scatter in each cell, and therefore more than one spectra per cell is needed to calibrate the photometric redshifts. On the other hand, increasing the grid size will lead to smaller average redshift scatter in a cell. For instance the average scatter value for a $120 \times 80$ grid is $\sim 0.02$). However, the larger the SOM grid is, trained by limited number of training data, the more interpolations are forced by the SOM. In the above mentioned grid of $120 \times 80$, this leads to $262$ cells ($\sim 3\%$) with no objects mapped to them and a minimum of $\sim 10 $k total spectra requirement to fill each cell with at least one spectra. The real LSST+\emph{WFIRST} sample would have orders of magnitude more galaxies compared to CANDELS and therefore all the cells will have sufficient data for the tomographic bins to have the cosmology required accuracy. However, making the SOM too large would require even more spectroscopic observations for calibration which is unnecessary given the already met requirement with a smaller SOM. 
 
\begin{figure*} 
\centering
  \includegraphics[trim=0cm 0cm 0cm 0cm, clip,width=1\textwidth] {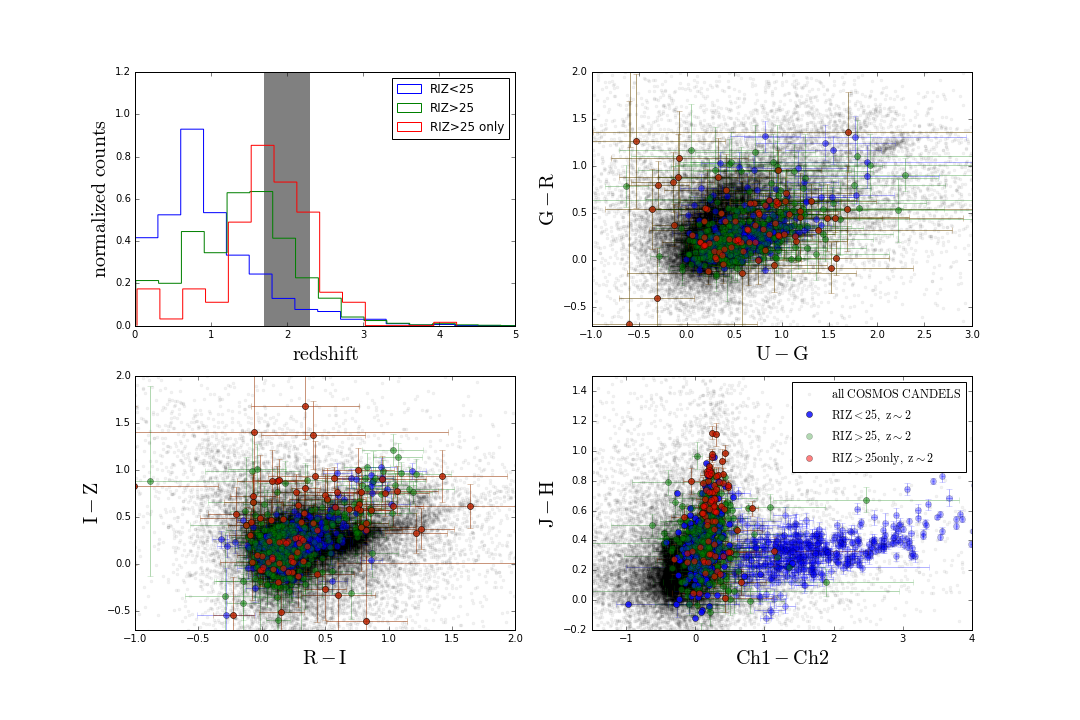}
\caption{Comparison of three subsamples in our \emph{WFIRST} lensing catalog: bright ($riz<25$; blue), faint ($riz>25$; green) and only faint ($riz>25$, and no bright SED in the object's SOM cell; red) galaxies in the same redshift range on color-color plots. Top-left: normalized distribution of subsamples are shown and the selected redshift range ($1.8<z<2.2$) for comparison in the other three subplots is shown as shaded gray region. $G-R$ vs. $U-G$ (top-right), $I-Z$ vs. $R-I$ (bottom-left), and $J-H$ vs. Spitzer/IRAC $Ch1-Ch2$ (bottom-right) plots show the distribution of all cosmos CANDELS galaxies with black and over plotted are the the three subsamples.}
\label{fig:faint_bright}
\end{figure*}

%\begin{figure*} 
%\centering
%  \includegraphics[trim=0cm 0cm 0cm 0cm, clip,width=1. \textwidth]%{IRAC_onSOM.png}
%\caption{Bright ($riz<25$- to \emph{Euclid} depth) and faint ($riz>25$ increased sample by WFIRST) samples mapped to the SOM and color-coded by IRAC Ch1-Ch2 colors on top ($z_{LLSST}-Ch1$; bottom) and the difference of the median color of faint and bright subsamples in the common cells on the right. Similar trends are seen among bright and faint galaxy samples in their IR colors with larger color differences occurring at cells with higher uncertainties.}
%\label{fig:IRAC_SOM}
%\end{figure*}

\begin{figure*} 
\centering
  \includegraphics[trim=0cm 0cm 0cm 0cm,width=1\textwidth] {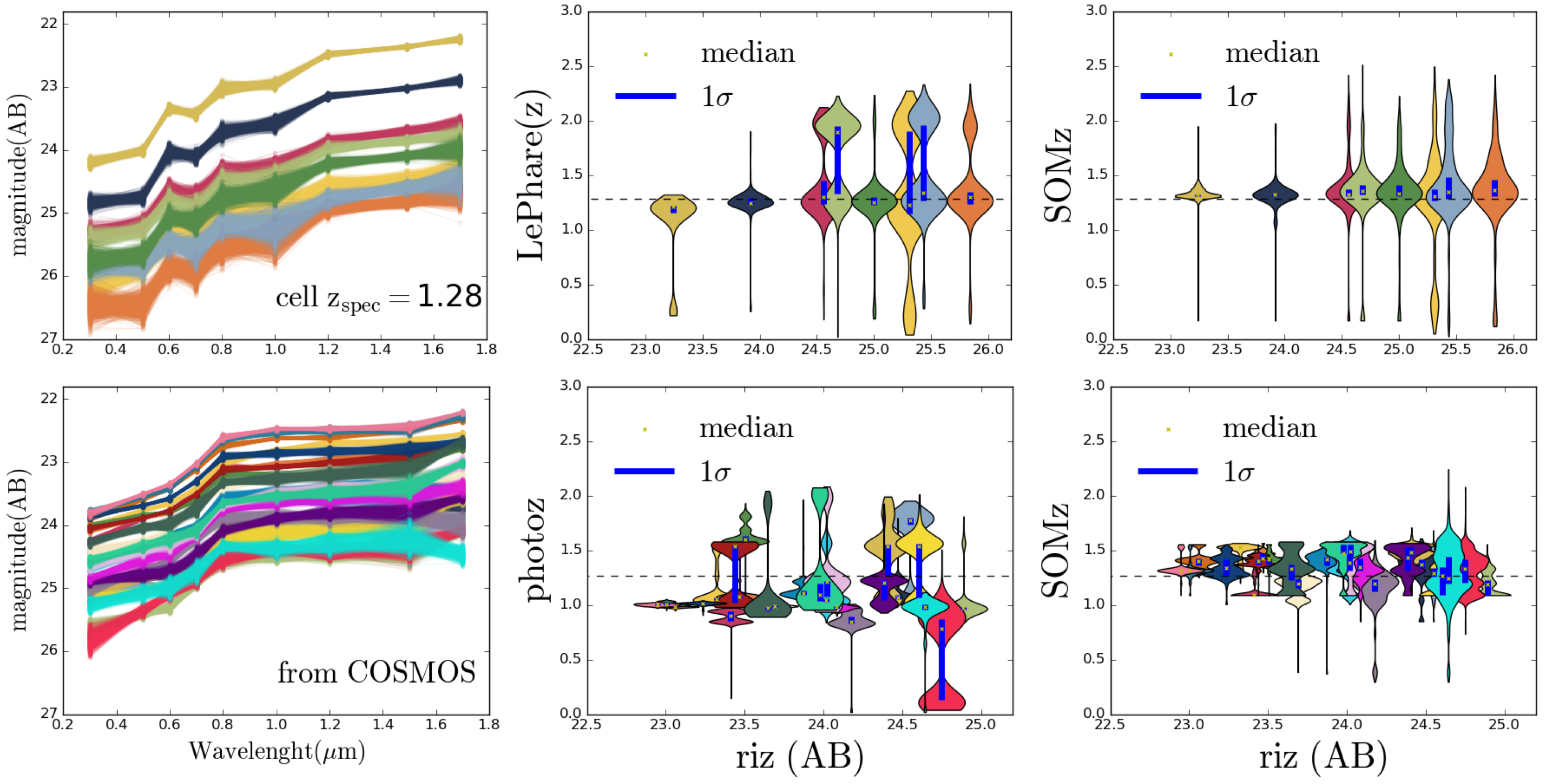}
\caption{Top-left: SED of seven galaxies with different brightnesses but similar SED shapes mapped to a cell at $specz=1.28$ are shown in different colors. From each SED we extract 1000 [Monte Carlo] realizations within the observed uncertainties (as seen from thickness of each SED). Larger spread of simulated SEDs in the fainter galaxies are due to larger photometric errors associated with them. Top-middle: photometric redshifts of each group of galaxies measured by the LePhare code colored by the colors of the SEDs in left panel. Blue vertical lines show one sigma spread around the median. The fraction of wrong redshift estimates get larger as one moves to fainter magnitudes due to larger uncertainties, as can be seen from the second peaks mostly at higher redshifts. Top-right: Same as middle panel with SOMz for each group of galaxies measured by mapping the simulated SEDs to the SOM. Redshift estimates get significantly better compared to the middle panel, with much smaller fraction of outliers. Bottom panels are the same as top, with 21 COSMOS galaxies mapped to the same cell of our trained LSST+WFIRST SOM.}
\label{fig:violin}
\end{figure*}

\subsection{The C3R2 survey}

 The C3R2 survey (\citealt{Masters2017}) is an ongoing spectroscopic effort to calibrate redshifts to the accuracy needed by weak lensing at the \emph{Euclid} depth by comprehensively mapping the empirical galaxy color-redshift relation. This is a good sample to at least partially fill the voids of the spectroscopy on the \emph{WFIRST} SOM as well. The C3R2 survey was initiated in 2015 with 10 allocated nights of Keck/DEIMOS observations (PI D. Stern) and 5 allocated nights of Keck/DEIMOS, MOSFIRE and LRIS (PI J. Cohen) followed by more nights on Keck, the VLT, and the GTC later. \cite{Masters2017} released the first season of observations, including 1283 high-quality redshifts and will release the second round of data in 2018 (Masters et al. in prep).

Unlike Euclid, which will use a broad optical (riz) filter to detect its weak lensing shear sample, the \emph{WFIRST} lensing sample will be derived from deep H band observations, leading to a sample that includes optically faint sources well beyond the \emph{Euclid} depth of $riz<25$. Given the depth achieved by C3R2, in the following sections we explore the extra spectroscopy needed to calibrate sources fainter than what would be studied by Euclid. 

\subsection{How different are faint and bright galaxies?}
 
The key question is whether the fainter galaxies comprising the \emph{WFIRST} lensing sample have different SEDs or if they lie in the same color-space as defined by Euclid. If the latter is the case, then the spectra acquired by the C3R2 to calibrate \emph{Euclid} would be sufficient to address the \emph{WFIRST} redshift calibration requirement as well. To first order approximation, Euclid's sample will contain galaxies with $riz<25$. To answer this question, we divide the sample into bright ($riz<25$) and fainter ($riz>25$) subsamples. In Figure \ref{fig:euclid}, we map the two subsamples onto the SOM separately (left panel: bright subsample, middle panel: faint subsample) and also identify cells which only contain faint galaxies (right panel). $\sim 95 \%$ of the SOM cells are covered by the bright sample, showing that fainter galaxies will not necessarily have different SEDs. While in the \emph{WFIRST} lensing sample, $\sim 26\%$ of objects have $riz>25$, they spread over $\sim 71 \%$ of the SOM, most of which also have bright galaxies mapped to them. The majority ($91\%$) of the faint sample ($riz>25$) live in color-cells which are also occupied by brighter \emph{Euclid} depth sources ($riz<25$). And only a small fraction of galaxies live in cells with no brighter counterpart ($\sim 4 \%$ of the SOM cells, $\sim 2 \%$ of the objects). 

We explore the cells with no bright object, and compare them to a neighboring cell. Figure \ref{fig:sed_faint} shows an example cell $\rm [8,25]$ in the SOM with only faint objects mapped to it in comparison with its neighboring cell $\rm [7,25]$ containing both faint and bright objects. As expected from the SOM (preserving topology), the SEDs of the two cells are very similar, and the difference in colors ($\sim 0.05$) are negligible. Note that, as the SOM is trained with colors and the exact magnitudes of each cell are not fixed, we manually assigned F184W magnitudes to the two SOM SEDs in Figure \ref{fig:sed_faint}, to plot the SEDs. The absence of bright objects in cells like these can then simply be due to small area coverage of CANDELS fields, rather than new or different types of galaxies.

 To examine this further, we compare the bright, faint and the faint with no similar bright galaxies samples across the same redshift range ($1.8<z<2.2$) on color-color plots in Figure \ref{fig:faint_bright}. The redshift range is fixed to eliminate the effect of distance on brightness. As can be seen in these color-color plots, fainter points with no similar bright SEDs (red data points) have much larger photometric errors compared to the bright objects (blue data points). As demonstrated before faint objects, do not occupy different portions of color space and have similar colors to the rest of the objects within their errors. Note that, in the lower right panel of Figure \ref{fig:faint_bright}, where Spitzer/IRAC Ch1-Ch2 color is plotted on the x axis, there is a distinct class of bright objects with no fainter counterparts (as per \citealt{Stern2005}, presumably AGNs), while all fainter objects do have neighboring bright object with similar colors. In short, fainter objects in our \emph{WFIRST} lensing sample live in the same color-space defined by brighter objects. 

%The SOM in this work, is trained using optical and near-IR colors covering observed $\sim 300-2000 nm$ (prerequisite of LSST+WFIRST). Therefore differences in the SEDs beyond this wavelength range will not be distinguishable. In Figure \ref{fig:IRAC_SOM}, we map the Spitzer/ IRAC $\rm Ch1-Ch2$ colors (top) and $\rm z_{LSST}-Ch1$ colors (bottom) of the bright ($riz<25$) and faint ($riz>25$) subsamples on to the SOM. The similarity between the overall trends of these colors among two subsamples suggests that even extending the wavelength regime to $\sim 5000 nm$, the faint and bright subsamples do not posses distinct SED shapes. 

\subsection{Redshift accuracy as a function of brightness}

We have demonstrated that the majority of faint and bright galaxies live in the same color-space, with similar SEDs. However, to be able to calibrate the \emph{WFIRST} sample with the C3R2 sample it is important to test for redshift accuracy as a function of brightness. Extra spectroscopy per cell is needed if the redshift scatter gets much larger as we include more faint galaxies to the SOM cell. 

In top panels of Figure \ref{fig:violin} we test for redshift accuracy as a function of brightness by simulating galaxies with similar SEDs at different brightnesses and measuring their redshifts. In this test, we chose a SOM cell with $specz=1.28$ having both faint and bright galaxies (4 $riz<25$ and 3 fainter than this threshold) from the lensing sample mapped to it. For each of these galaxy we generated 1000 similar SEDs within their uncertainties at each band. We measure the redshift of each simulated SED using the SED fitting code \textsc{LePhare} (\citealt{Arnouts1999}, \citealt{Ilbert2006}) as well as using the SOM (i.e. SOMz). The middle panel of Figure \ref{fig:violin} shows the distribution of photometric redshifts for simulated galaxies in different bins of riz magnitude. This panel shows that moving to fainter magnitudes, the fraction of wrong redshift assignments gets significantly larger, as seen from the second peak in the distribution. The blue vertical lines on the violin plots, representing the 1 sigma value of redshift distributions, are larger than allowed for redshift calibration in each cell. Therefore, SED fitting codes will not be able to provide the redshift uncertainty required for the fainter subsamples. However, as we show in the right panel of Figure \ref{fig:violin}, if the color-redshift relation from the SOM is used to measure the redshifts (SOMz), the scatter significantly decreases. SOMz is measured by mapping the simulated galaxies to the SOM. As can be seen from this figure, there is no bias in the median redshifts with brightness when using the SOM method and the dispersions are also of the order ($\sim 0.04$) allowed by the SOM calibration technique. 

 We note that in the SOMz method here, the mapping of color to redshift is generated from the median CANDELS photoz of galaxies in each cell. Ideally, once the SOM is covered with spectroscopic redshifts, the mapping would be more accurate and less dependent on the prior use of SED fitting codes as the case here. To assure that the improvement we see by using the SOMz method is not fully based on the prior use of the same galaxies in training the SOM and making the color-redshift relation, we extend our test (bottom panels of Figure\ref{fig:violin})to galaxies from the COSMOS catalog (\citealt{Laigle2016}). We used the closest 9 photometric bands to generate and map the colors to our SOM. We measured the photoz and SOMz of thousand realizations of the 21 galaxies which map to that same cell on the SOM. While these galaxies are all relatively bright ($riz<25$) and there are slight discrepancies in the photometries used (e.g., Ultravista near IR photometries in COSMOS vs. HST in CANDELS), the improvement from traditional SED fitting is still evident.

\subsection{Spectroscopy Recommendation}

As shown in the previous sections, spectroscopic sampling by the C3R2 survey for the Euclid mission is sufficient for filling more than $90\%$ of the color-space with at least one spectra. To fill out the remainder of the \emph{WFIRST} SOM, $200$ new spectroscopic redshifts are needed to fill the color-space with at least one spectra at each cell. This corresponds to the $4\%$ of the cells that have only faint objects associated with them. With $1.2$ k spectroscopy, cells with larger dispersions ($25\%$ of cells where redshift scatter $>0.05$) will have two spectroscopic redshifts mapped to them. This would be helpful in cells where most of their associated galaxies are faint. Therefore, we recommend $\sim 0.2-1.2$ k extra spectra, to fully calibrate the photometric redshifts of LSST+\emph{WFIRST} lensing sample.

In addition to weak lensing calibration, from the galaxy evolution point of view, it is absolutely important to obtain spectroscopic observations of these faint systems which have not been spectroscopically observed before. This will enable comparison of their more detailed physics to the brighter galaxies with similar broad band SEDs (colors).

The number of recommended spectra needed to calibrate the LSST+\emph{WFIRST} color-redshift relation is not large. Yet, spectroscopy of these faint targets would not be easy. Most of the voids in previous spectroscopic observations that we found by using the SOM, are likely due to the biased selection techniques. Another possibility however can be due to unsuccessful spectroscopic observations, i.e. non-sufficient observing time or wrong telescope/instrument chosen for the observations. Most of the cells with no spectroscopic coverage are in the $z\sim 1-2$ regime, once dubbed the redshift desert due to historical difficulties. Powerful instruments on ground and space-based telescopes such as the Keck twin telescopes, the future Thirty Meter Telescope (TMT), and the James Webb Space Telescope (JWST) can now/will explore this redshift regime and to fainter depths. 

%%%%%%%%%%%%%%%%%%%%%%%%%%%%%%%%%%%%%%%%%%%%%%%%%%%%%%%%%%%%%%%%%%%%%%%%%%%%%%%%%%%%%%%%%%%%%%%%%%%%%%%%%%%%%%
\section{Summary \label{sec:disc}}

In this paper, we studied the redshift calibration requirements for \emph{WFIRST} HLS weak lensing analysis to meet the stage IV dark energy desired accuracy. We adopted the methodology introduced by M15, which calibrates the color-redshift relation using SOMs. We imitated the LSST+\emph{WFIRST} lensing sample using optical and near-IR data from the five CANDELS fields and trained a SOM with successive colors of galaxies in the LSST+\emph{WFIRST} filter set. The smoothness of redshift distribution on a SOM trained by colors, illustrates the color-redshift relation and makes the SOM an optimal source for spectroscopic target selection. 

Based on Monte Carlo simulations in M15, and given the estimated average redshift uncertainties in our SOM cells, a tomographic bin containing $\sim 200$ SOM cells, would be sufficient to reach $\Delta \langle z \rangle / (1+\langle z \rangle) < 0.002 $. For the technique to be efficient, most SOM cells need to have at least one spectroscopic object mapped to them for calibration. This is equivalent to $\sim 5 $k total spectroscopic redshifts to calibrate the \emph{WFIRST} SOM. However, in addition to the already existing spectroscopic observations in the CANDELS fields (covering $57\%$ of the SOM cells) the C3R2 survey is filling the color-space of \emph{Euclid} galaxies with spectroscopic observations.  

\begin{figure} 
\centering
  \includegraphics[trim=0cm 0cm 0cm 0cm, clip,width=0.49 \textwidth] {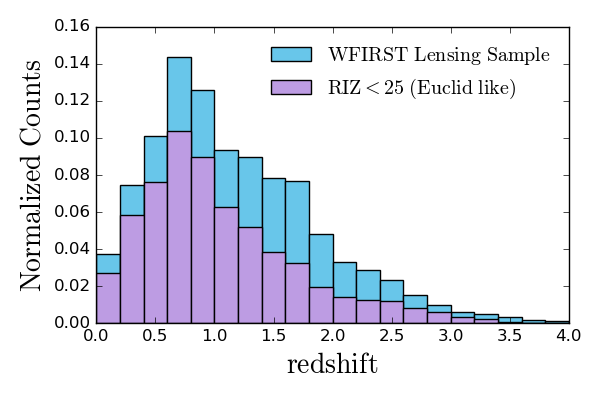}
\caption{\emph{WFIRST} will significantly increase the lensing sample size. Normalized redshift distribution of galaxies in the LSST+\emph{WFIRST} analog sample is shown as the blue histogram and the fraction of galaxies with $riz<25$ as would be found by \emph{Euclid} is over plotted in purple.}
\label{fig:redshifts}
\end{figure}

We showed that $\sim 26\%$ of the \emph{WFIRST} lensing sample consists of sources fainter than the \emph{Euclid} depth in the optical, $91\%$ of which live in color cells also occupied by brighter (\emph{Euclid}-depth) sources. We demonstrated the similarity between the fainter and brighter subsamples in same cells as well as the feasibility of measuring the redshifts of fainter objects to the accuracy needed using the SOM color-redshift relation. Since the $\sim 4\%$ of cells which have only fainter objects associated to them, might be due to small sample size in the CANDELS fields as well as larger photometric errors in the fainter sample, we recommend extra spectroscopy for these cells to calibrate the color-redshift relation on \emph{WFIRST} SOM throughly. We recommend $\sim 0.2-1.2 $k new spectra, which will cover the cells with only faint objects as well as those with large redshift dispersions. It is crucial to note that, having most of the calibration already in place by the C3R2 \emph{Euclid} effort does not imply similarity between \emph{WFIRST} weak lensing cosmology and the less deep surveys, as the lensing sample size will increase significantly (see Figure \ref{fig:redshifts}).

In our analysis, we have used an interpolation technique to estimate photometry in LSST+\emph{WFIRST} filter sets based on available photometry in different filter sets. The method is tested extensively and is the easiest logical way to reproduce statistically correct distributions of photometries and colors of galaxies. Future works using improved techniques will test for the robustness of the interpolation on an object by object basis. 

One weakness of this work is the small sample size used for training the SOM. As discussed in the paper, CANDELS data were the best available option due to comparable depth of observations to those expected from the \emph{WFIRST} HLS. CANDELS observations being done in five well-separated fields in the sky should mitigate the effect of cosmic variance. We found the color-space of our \emph{WFIRST} lensing sample to be representative of a sample trained for Euclid (19 times larger in area), which suggests that the effect of cosmic variance in the SOM calibration technique should be minimal. However, very large area simulations are needed to enable a more quantitative investigation of this effect on the findings presented in this work (Capak et al. in prep.). We will revisit our forecasts once \emph{WFIRST} observations are available, and retrain a SOM with actual observations for photometric redshift calibration as well as selection of weak lensing tomographic bins.

\acknowledgements
We wish to thank the anonymous referee for constructive comments that significantly improved this work. S.H. is greatful to Chris Hirata for constructive comments and for providing the lensing sample selection criteria. S.H. also thanks Rebecca Larson for carefully reading the manuscript. This work used {\sc SOMPY}, a python package for self organizing maps (main contributers: Vahid Moosavi @sevamoo, Sebastian Packmann @sebastiandev, Iv\'an Vall\'as @ivallesp). This research made extensive use of data from the CANDELS survey. Parts of this research were carried out at the Jet Propulsion Laboratory, California Institute of Technology, under a contract with the National Aeronautics and Space Administration.

%%%%%%%%%%%%%%%%%%%%%%%%%%%%%%%%%%%%%%%%%%%%%%%%%%%%%%%%%%%%%%%%%%%%%%%%%%%%%%%%%%%%%%%%%%%%%%%%%%%%%%%%%%%%%%

%\bibliography{wps.bib}

\end{document}